\documentclass[10pt]{article}

\usepackage{booktabs} 

\usepackage[ruled]{algorithm2e} 

\SetAlFnt{\small}
\SetAlCapFnt{\small}
\SetAlCapNameFnt{\small}
\SetAlCapHSkip{0pt}
\IncMargin{-\parindent}

\usepackage{url}
\usepackage{amsfonts}

\usepackage{xcolor}
\usepackage{tcolorbox}

\usepackage{geometry}

\newcommand{\code}[1]{\texttt{\small #1}}
\newcommand{\snapeLong}{Prof. Ivano Salvo}
\newcommand{\snape}{Prof. Salvo}
\newcommand{\hermioneLong}{Ms. Agnese Pacifico}
\newcommand{\hermione}{Agnese}
\newcommand{\mrsGranger}{Ms. Pacifico}
\newcommand{\studenteUno}{Elena}
\newcommand{\studenteDue}{Dario}
\newcommand{\myParagraph}[1]{\smallskip\emph{#1}}
\usepackage[inline]{enumitem}
\usepackage{graphicx}

\title
{Computing Integer Sequences: Filtering vs Generation
\\(Functional Pearl)}

\author{Ivano Salvo$^{\dag\ast}$, Agnese~Pacifico$^\ddag$ \medskip
\\
$^\dag$ Departement of Computer Science, Sapienza University of Rome\\
$^\ddag$ Departement of Mathematics, Sapienza University of Rome\medskip
\\
$^\ast$ Corresponding author: \texttt{salvo@di.uniroma1.it}
}
\date{}

\begin{document}
\maketitle

%

\begin{abstract}
As a report of a teaching experience, we analyse Haskell programs computing 
two integer sequences: the 
Hamming sequence 
and 
the Ulam sequence. 
For both of them, we investigate two strategies of computation: 
the first is based on filtering out those natural numbers that do not belong to the sequence, 
whereas the second is based on the direct generation of numbers that belong to the sequence.
Advocating cross-fertilisation among ideas emerging when programming in 
different programming paradigms, 
in the background, we sketch out some considerations 
about corresponding C programs solving the same two problems.
\end{abstract}
\maketitle


\section{Prologue}

\myParagraph{\snapeLong\ (writing to a colleague):}
[\ldots] Teaching can be very boring. 
Starting to teach a new course, the first year, one decides what to do according to the course program.
The second year, one tries to arrange things that did not work so well  
during the previous year, 
probably taking into account comments in student evaluation forms, 
or working on students' lack of preparation 
as highlighted by exams. 
The third year, one makes it perfect. At least from his or her point of view. 
And then\ldots? How one can save himself, and of course his students, from tedium?

I think that you will agree with me that one possible therapy is 
transforming teaching into learning and/or ``research''. 
Here, ``research'' can mean ``find new solutions to old problems'', 
or just ``find some amusing new examples'', sometimes just searching throughout books or the web
and adapt or re-think their solution for your course.
. In any case, every year I prefer to choose a theme, a leitmotif that drives my research of new topics [\ldots]. 

\myParagraph{\snape\ (first lecture, course presentation):}
Dear pupils, welcome to the course {\sl ``Functional and Imperative Programming Pearls''}. %
In this course, we will introduce C and Haskell core languages,  
and we will discuss imperative and functional programming styles,  
by comparing several solutions to programming problems solved in both these two languages.

This semester, I will get inspiration from the work of Edsger W. Dijkstra. 
Many intriguing programming examples presented during lectures 
and/or proposed as homework assignments will be chosen from   
Dijkstra's classic book ``A Discipline of Programming'' \cite{dijkstra76} 
and from Dijkstra's ``digitalised'' manuscripts \cite{dijkstra-web}.  

Exactly twenty years ago, 
when I was a first year Ph.D. student, I attended a 
computer science summer school on Formal Methods for Software Engineering. 
In one course of that school, prof. Dijkstra presented some brilliant solutions 
to small (but not trivial at all) programming exercises. 
I can remember quite well only one of them, 
about determining whether two arrays are equal modulo shift 
(and of course, as soon as I became a teacher, I included (an adaption of) it 
as optional material in my programming courses), 
but I can never forget the magic atmosphere of prof. Dijkstra's lectures.

At the beginning of each lecture, he would just put a blank slide on the projector 
(at that time Power Point was already very popular, but prof. Dijkstra was 
a tireless amanuensis) 
and start to write the post-conditions  
that the program should satisfy at the bottom of the slide.  
Then he started writing commands 
(obviously in Dijkstra's guarded command language) 
mixed with logical assertions backward from the bottom to the top, until   
(just below the name of the function)   
function reasonable pre-conditions would appear.

\section{The Hamming Problem}
 
\myParagraph{\snape:} 
In {\sl ``A Discipline of Programming''} (\cite{dijkstra76}, Chapter 17, ``An Exercise Attributed to R.~W.~Hamming''), prof. Dijkstra proposed the following problem:
\begin{quote}
{\em
``To generate in increasing order the sequence 1, 2, 3, 4, 5, 6, 8, 9, 10, 12, \ldots of 
all numbers divisible by no primes other than 2, 3, or 5. Another way of stating which values are in the sequence is by means of three axioms:
\begin{enumerate}
\item
\label{ax1}
1 is in the sequence.
\item
\label{ax2}
If $x$ is in the sequence, so are $2x$, $3x$, and $5x$.
\item
The sequence contains no other values than those that belong to it on account of \ref{ax1} 
and \ref{ax2}.
\end{enumerate}
\smallskip

[\ldots] Being interested only in terminating programs, we shall make a program generating only the, say, first 1000 values of the sequence.''
}
\end{quote}

Working in Haskell, the last sentence looks funny: thanks to lazy evaluation, 
we can ``easily'' write a program manipulating infinite data structures and, in this case, we can write a program that generates the infinite list representing the desired sequence. 

To make this exercise more intriguing,  
I want you to solve a generalisation of this problem, so I ask you to 
write a program that generates 
``all numbers divisible by no primes other than those in a given set 
$P$ of primes''. 
We call $C_P$ the ordered sequence of numbers that are product of 0 or more 
primes in $P$. 
Treating $C_P$ as a set for convenience, 
we can 
concisely define $C_P$
as the smallest set such that 
$1\in C_P$, and for all $c\in C_P$ and for all $p\in P$ we have that $c\cdot p\in C_P$. 
In other words, $C_P$ is simply the smallest set closed under products of primes in $P$.

\subsection{A filter program for the Hamming problem}\label{sec:hamming}
\myParagraph{\hermioneLong:}
I have already written an easy Haskell program 
during your explanations of the problem  (Fig.~\ref{fig:composti-filter}): 
this program filters numbers that cannot be factorised by using only numbers in \code{p} out of the list of all natural numbers. 

Function \code{compositeOf p n} adapts a function I wrote as a solution to a past homework: it checks if \code{n} is a composite of the numbers in the list \code{p} only. By using the higher order function \code{filter}, we can write a one line Haskell definition of the (generalised) Hamming sequence.

\myParagraph{\snape:}
Unfortunately, performances of your program are very poor. 
The fact is that composites of a finite set of natural numbers 
are quite sparse inside natural numbers. 
This is immediately evident if we consider the set $P=\{2\}$: in this case, $C_P$ is the sequence of powers of 2. Only a really dummy programmer would compute powers of two by checking all natural numbers, verifying one by one if they can be factorised by using $2$ only. 

The same holds for many other integer sequences, such as Fibonacci numbers and, I suppose, for all sequences with density 0 and a reasonably simple generating rule.

Calling $A(n)$ the number of terms of a sequence smaller than $n$, the density of a sequence is defined as $\lim_{n\rightarrow\infty} \frac{A(n)}{n}$.
For example, the number $A(n)$ for {\em regular numbers}, as mathematicians call composites of $2$, $3$, and $5$, is estimated as $\frac{\log_2 n \log_3 n \log_5 n}{6}$ and hence Hamming numbers' density is 0. Intuitively, this is true for any finite set of primes $P$.

Even though you correctly follow the Haskell idiom to name a list pattern \code{x:xs}, I prefer 
(probably because I'm an ML native speaker) to adhere to the following name convention: if \code{x} is the name of a list, I call \code{hx} its head (`{\em h}ead of \code{x}') and \code{tx} its tail (`{\em t}ail of \code{x}'). 
I find it useful to use patterns of the form \code{x@(hx:tx)} whenever the whole list \code{x} is needed in the right hand side of a definition. 
This is, of course, a matter of taste, and an exception with respect to my usual suggestion to follow consolidated language idioms.

%
%

\begin{figure}
\begin{center}
\begin{minipage}{0.68\textwidth}
\begin{tcolorbox}[colback=white, boxrule=0pt, toprule=1pt, bottomrule=1pt, left=0pt]
{\small
\begin{verbatim}
   compositeOf     p     1  = True
   compositeOf     []    n  = False
   compositeOf (x:xs) n 
       | mod n x == 0 = compositeOf (x:xs) (div n x) 
       | otherwise    = compositeOf xs n 

   cp p = filter (compositeOf p) [1..] 
\end{verbatim}
}
\end{tcolorbox}
\end{minipage}
\end{center}
\vspace{-3mm}
\caption{A filter solution to the Hamming problem.}
\label{fig:composti-filter}
\vspace{-3mm}
\end{figure}

\subsection{Generative solutions to the Hamming problem}
\myParagraph{\snape:} The solution suggested by prof. Dijkstra,  
of course derived by analysing logical assertions, 
and essentially based on monotonicity of the product function, 
is the following:  
having computed the first $n$ numbers in the sequence, let's say $c_1,c_2, \ldots, c_n$, 
the number $c_{n+1}$ can be computed as the minimum of 
$\{2c_i\ |\ 1\leq i\leq n \ \& \ 2c_i>c_n\}\cup\{3c_i\ |\ 1\leq i\leq n\ \&\ 3c_i>c_n\}\cup\{5c_i\ |\ 1\leq i\leq n\ \&\ 5c_i>c_n\}$. 
To find $c_{n+1}$ however, by monotonicity, it is enough to compute 
the minimum of the set $\{2c_{i+1}, 3c_{j+1}, 5c_{k+1}\}$ where 
$2c_i$ (and $3c_j, 5c_k$ respectively) is the last composite of 2 (and 3, 5 respectively) 
already inserted in the sequence. 
And, more or less, the same line of reasoning I expected from you, in  
your C programs of your homework.
 
The main obstruction in writing a Haskell program that applies this idea 
is understanding how to deal with the infinite list under construction and how to 
``record'' the information about the indices $i,j,k$ 
of the last multiples of $2$, $3$, and $5$ inserted in the sequence.
The first issue is quite easy, once you have understood 
some basic and classical examples of lazy programs, such as Fibonacci numbers or the Sieve of Eratosthenes.  
At first glance, the second issue is a bit more complex, 
but again, the Fibonacci example can provide some useful hints to you.
 
\myParagraph{\hermione:} 
I did it! To simplify the problem, I started by considering the case in which the set 
$P$ has just two elements, let's say $P=\{x,y\}$. 
I have defined a four parameters function \code{nextC x y cx cy} 
(Fig.~\ref{fig:composti-generative}), 
where \code{x} and \code{y} are the numbers in $P$, and \code{cx} and \code{cy} 
are two tails of the (infinite) 
list \code{cp} under construction representing $C_P$. 
The first element of \code{cx} (i.e. \code{hcx}), 
is the first element of \code{cp} such that \code{fx=x*hcx} 
has not yet been added so far to \code{cp} 
and, similarly, the first element of \code{cy} (i.e. \code{hcy}), 
is the first element of \code{cp} such that \code{fy=y*hcy} 
has not yet been added so far to \code{cp}.
Therefore, \code{fx} and \code{fy} are the next candidates 
to be inserted in the sequence of composites. We choose the minimum between 
them and we make a step forward on the corresponding list in the recursive 
call of \code{nextH}. Of course, \code{fx} can be equal to \code{fy}. In this case,  
we can insert indifferently one of them and make a step forward on both lists.

\myParagraph{\snape:} 
Along the same lines, you can solve the classical version of the Hamming problem with $P=\{2,3,5\}$ 
by defining a function of six parameters that, at each recursive call, 
computes the minimum among three numbers and  
distinguishes all combinations of numbers equal to such minimum\ldots 
but its code would be unnecessarily cumbersome. 
It would be much better to approach the general problem, assuming the set $P$ has   
an arbitrary cardinality.

\myParagraph{\hermione:}
First of all, it is clear that parameters \code{x} and \code{y} 
are useless. Laziness allows us to define the sequence of composites of 
\code{x} as \code{map (x*) hns} (the name \code{hns} stands for {\em H}amming {\em n}umber{\em s}).
Therefore, we can consider the list of infinite lists of composites, 
one for each element in $P$, that can be computed as 
\code{map (\textbackslash x->(map (x*) hns) p}.
We can: \begin{enumerate*}[label=({\arabic*})]
\item 
\label{hamming:one}
compute the next element to add to \code{hns} as the minimum \code{m} of the 
finite list of heads of such lists, and 
\item 
\label{hamming:two}
make a step forward on all those lists whose 
head is \code{m}. 
\end{enumerate*}

As for \ref{hamming:one}, we can select the list of list heads with 
\code{(map (\textbackslash z-> (head z)) c)} 
and computing its minimum by properly folding the \code{min} function\ldots

\begin{figure}
\begin{center}
\begin{minipage}{0.68\linewidth}
\begin{tcolorbox}[colback=white, boxrule=0pt, toprule=1pt, bottomrule=1pt, left=0pt]
{\small
\begin{verbatim}
  nextH :: (Ord t, Num t) => t -> t -> [t] -> [t] -> [t]

  nextH x y cx@(hcx:tcx) cy@(hcy:tcy)
      | fx  < fy  = fx:nextH x y tcx  cy
      | fx  > fy  = fy:nextH x y  cx tcy
      | otherwise = fx:nextH x y tcx tcy
     where  
         fx = x * hcx
         fy = y * hcy

  cp = 1:nextH 2 3 cp cp
\end{verbatim}
}
\end{tcolorbox}
\end{minipage}
\end{center}
\vspace{-3mm}
\caption{A generative solution to the Hamming problem. Two numbers version.}
\label{fig:composti-generative}
\vspace{-3mm}
\end{figure}

\myParagraph{\snape:} The $\eta$--rule, \mrsGranger! Do not forget the $\eta$--rule from the $\lambda$-calculus, \mrsGranger. 

\myParagraph{\hermione:}
Right! I'm sorry\ldots As for \ref{hamming:two}, we remove all list heads that are equal to the just computed new Hamming number, 
again by using the  
higher order function \code{map} 
(Fig.~\ref{fig:composti-generative-2}).  

\myParagraph{Other Student:} Are you confident that no productivity problems arise in your program?

\myParagraph{\hermione:} Yes, because function \code{nextHG} at each invocation computes {\em always} a new element of 
the resulting infinite list \code{cp} and consumes {\em at most} one element of each list in the list of lists \code{c}. 
Therefore, the head of all lists in 
\code{c} of Hamming candidates is always computed {\em before} it is needed in \code{nextHG}. 

\myParagraph{\snape} Right. Analysing your program, by selecting the minimum of heads of a list of lists, 
you have implicitly generalised the \code{merge} function to an arbitrary finite number of lists. 
Maybe better to call it \code{union}, since we need to avoid duplicates in the resulting list. 
Essentially, you have just re-discovered the classic functional solution to the Hamming problem 
(with numbers 2, 3, 5) in the Bird-Wadler book \cite{birdWadler88}.

\begin{figure}
\begin{center}
\begin{minipage}{0.68\linewidth}
\begin{tcolorbox}[colback=white, boxrule=0pt, toprule=1pt, bottomrule=1pt, left=0pt]
{\small
\begin{verbatim}
  nextHG :: Ord a => [[a]] -> [a]

  nextHG c = m:nextHG y where
     m = foldr1 min (map head c)
     y = map (\x@(hx:tx)->if hx==m then tx else x) c

  cp p = hns
     where hns = 1:nextHG (map (\x-> map (x*) hns) p)
\end{verbatim}
}
\end{tcolorbox}
\end{minipage}
\end{center}
\vspace{-3mm}
\caption{A generative solution to the Hamming problem. General version I.}
\label{fig:composti-generative-2}
\vspace{-3mm}
\end{figure}


\begin{figure}[b]
\begin{center}
\begin{minipage}{0.68\linewidth}
\begin{tcolorbox}[colback=white, boxrule=0pt, toprule=1pt, bottomrule=1pt, left=0pt]
{\small
\begin{verbatim}
  union :: Ord a => [a] -> [a] -> [a]

  union x@(hx:tx) y@(hy:ty)
     | hx == hy = hx:union tx ty
     | hx <  hy = hx:union tx y
     | hx >  hy = hy:union x  ty 

  cp p = hns
       where hns = 1:foldr1 union (map (\x-> map (x*) hns) p)  
\end{verbatim}
}
\end{tcolorbox}
\end{minipage}
\end{center}
\vspace{-4mm}
\caption{A generative solution to the Hamming problem. General version II.}
\label{fig:composti-generative-union}
\vspace{-2mm}
\end{figure}
       
\myParagraph{\hermione\ (searching frenetically in a book):} Oh... yes. A good idea could be to make this explicit,  
by defining the \code{union} of an arbitrary number of lists, by folding the binary \code{union} function 
inside a list of ordered lists (Fig.~\ref{fig:composti-generative-union}). This, essentially, solves the Exercise 7.6.5 in \cite{birdWadler88} and works also if \code{p} contains non-prime numbers.

%
%


\myParagraph{\snape:} 
Excellent, \mrsGranger. To be honest, I do not know anything about 
prof. Dijkstra's attitude towards recursive programs and 
functional programming. After all, all problems in his book are 
solved by iterative programs\ldots

\myParagraph{\hermione:} I have found some documents on this topic! At the beginning, he was rather skeptical 
about functional programming. For example, he firmly criticised John Backus Turing Award lecture~\cite{Backus78} in one of his 
writings~\cite{dijkstraOnBackus}. But some years later he supported Haskell against Java~\cite{dijkstraOnHvJ}
as a good choice as a programming language in an introductory programming course. Notwithstanding this, 
Java was chosen instead of Haskell all the same! 
 
\myParagraph{\snape:} 
Nevertheless, I think he would have appreciated the elegance of these programs. 
It is hard to think of an imperative program solving the same problem 
as much small and elegant as them. 
Their ``holistic'' style in manipulating an infinite list 
(and a list of infinite lists) as a whole is perfectly 
in agreement with the functional style promoted by John Backus in his seminal work \cite{Backus78} 
and beloved by functional programmers.  

%

\section{Ulam Numbers}
\label{sec:ulamNumbers}
\myParagraph{\snape:} 
Also in teaching activity, research often means leaving known and prefigured paths. 
 Until now, I have proposed to you several problems from Dijkstra's writings. 
However, 
a sentence written by prof. Dijkstra in the presentation of the Hamming problem (\cite{dijkstra76}, Chapter 17):
\begin{quote}
{\em
``We include this exercise because its structure is quite typical for a large 
class of problems.''
}
\end{quote}
led me to look for some other intriguing integer sequences to be computed.
My attention has been captured by Ulam numbers \cite{ulam64}.

The Ulam sequence is defined as follows: 
$u_1=1, u_2=2$ and $u_{n+1}$ is the least strictly greater than $u_n$ number that is a unique sum of two distinct earlier terms.
It defines the following infinite sequence (here all terms up to 100):
\[
\begin{array}{c}
1, 2, 3, 4, 6, 8, 11, 13, 16, 18, 26, 28, 36, 38, 47, 48, 53, 57, 62, 69, 72, 77, 82, 87, 97, 99,\ldots
\end{array}
\]
For a quick summary of properties (and mainly unanswered questions 
and unproven conjectures) of this sequence, you can refer to the Ulam page in {\em The 
On-line Encyclopedia of Integer Sequences} (OEIS) \cite{ulam-oeis}.
Just to  quickly understand this sequence, 5 is not in the sequence because 
$5=3+2=4+1$ and $1, 2, 3, 4$ are all already in the sequence: therefore 5 can be obtained 
as two distinct sums of distinct earlier terms. 
There are also numbers that are not in the sequence because no pair of earlier numbers gives them as their sum. These numbers form themselves an infinite 
sequence \cite{not-ulam-oeis}, let's call it $v$, that starts with the number $v_1=23$ (here all terms up to 100):
$$23, 25, 33, 35, 43, 45, 67, 92, 94, 96,\ldots$$

Despite being quite rare up to 100, they are ``more'' than Ulam numbers. 
For example, up to $u_{1000}=12294$, we have $2173$ numbers that are not sums of two Ulam numbers, and $v_{2173}=12287$.

Computing the fascinating Ulam sequence has received attention recently
\cite{gibbsMcCranie,knuth16}: motivated by computing up to trillions of terms 
in reasonable time, these works investigate 
heuristics relying on 
some unproven  properties (but checked experimentally) of Ulam numbers distribution.
These works are far from our main goal here, which is writing programs based on 
``direct'' methods, i.e. just by applying definitions and some immediate optimisations.

As the definition of the Ulam sequence  immediately suggests, 
Ulam numbers can be computed by following an approach based on filtering. 
Indeed, having $u_1, u_2, \ldots, u_n$ in your hands, 
it is intuitive to check if $u_n+1$ is $u_{n+1}$, by checking 
how many distinct sums of pairs of distinct numbers  
among $u_1, u_2, \ldots, u_n$ give 
as result $u_n+1$.
If $u_n+1$ cannot be obtained as a sum of earlier terms, or 
it can be obtained by more than one pair of earlier terms, 
then we try with $u_n+2$ and we have to continue until we find $u_{n+1}$ that can be 
at worst $u_n+u_{n-1}$: this sum is strictly greater than any other sum that 
we can get from a pair of numbers in the increasing sequence $u_1, u_2, \ldots, u_n$.  Incidentally, I observe here that 
this argument proves that Ulam numbers form an infinite sequence. 
You should prepare as Homework, a C function: 
\code{int nextU(listDL L);}  
that, assuming as a precondition that the doubly linked list \code{L} contains the first $n$ Ulam numbers, returns as a result $u_{n+1}$, 
adding at the same time $u_{n+1}$ at the end of \code{L}. 

We will discuss Haskell solutions to the Ulam numbers computation in the next lecture.

\subsection{Searching the WEB}
\myParagraph{\hermione:}
I have found on the WEB this 
one-line horrific code \cite{haskellUlam67} of 67 characters only (Fig.~\ref{fig:g1}):

\begin{figure}
\centering
\vspace{-1mm}
\begin{minipage}{.68\textwidth}
\begin{tcolorbox}[colback=white, boxrule=.7pt, toprule=.7pt, bottomrule=.7pt, left=-1pt, bottom=20pt]
\centering
\includegraphics[scale=0.424]{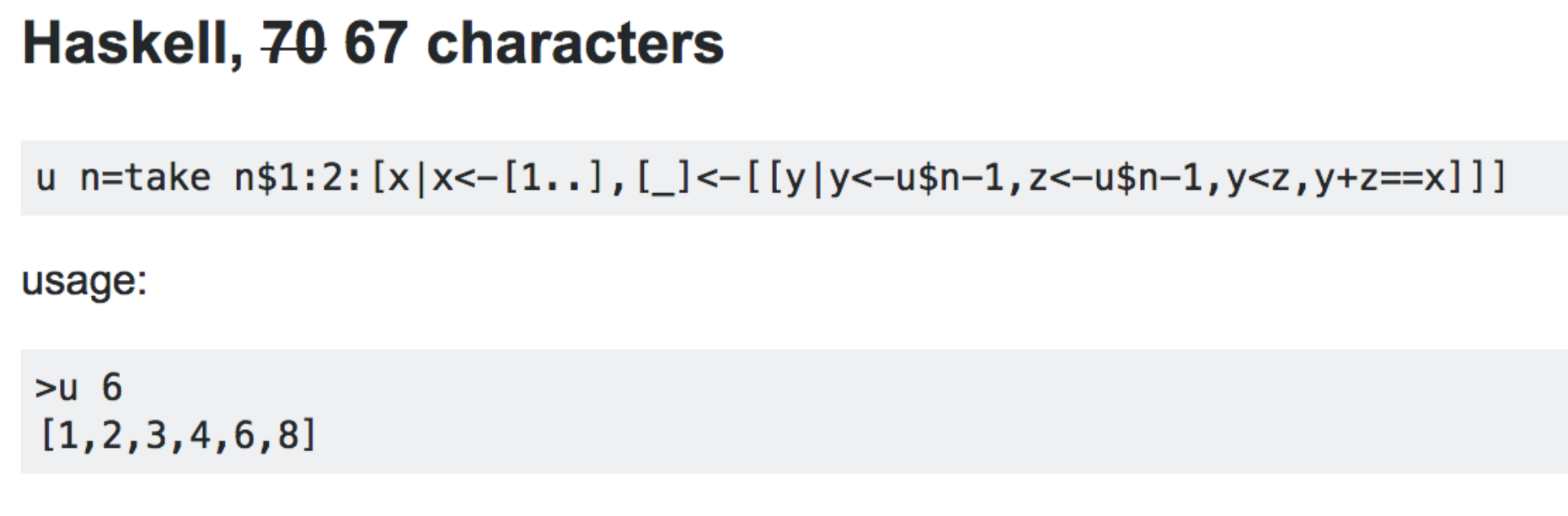}
\vspace{-8mm}
\end{tcolorbox}
\caption{Ulam program of 67 characters from Code Golf \cite{haskellUlam67}}
\label{fig:g1}
\end{minipage}
\vspace{-2mm}
\end{figure}



\myParagraph{\studenteUno:}
This abnormal program belongs to the Code Golf community, that is those 
Haskell, C, or other language programmers 
that love challenges like writing a program with the minimum number of characters.  
Other virtuosities include, for example, self-reproducing code.  

\myParagraph{\snape:}
Any interested student is invited to understand what this program exactly does\ldots

\myParagraph{\studenteDue:}
Roughly speaking, it selects Ulam numbers from the sequence of natural numbers, 
by selecting those numbers $x$ such that the list of $y$ such that there exists $z$ such that $x=y+z$  
is a list of just one element 
(of course, $y$ and $z$ are taken from the list of already computed Ulam numbers).
Unfortunately, this program is so slow that even computing $u_8$ takes several seconds on the \code{GHCi} Haskell interpreter\ldots

\myParagraph{\snape:}
So, for exercise, try to prove its termination. I'm not sure it terminates for $n>9$.
Some virtuoso Haskell programmer can try to write an equivalent program of 66 characters or less!

\begin{figure}
\centering
\vspace{-3mm}
\begin{tcolorbox}[colback=white, boxrule=.7pt, toprule=.7pt, bottomrule=.7pt, left=-1pt]
\includegraphics[scale=0.5]{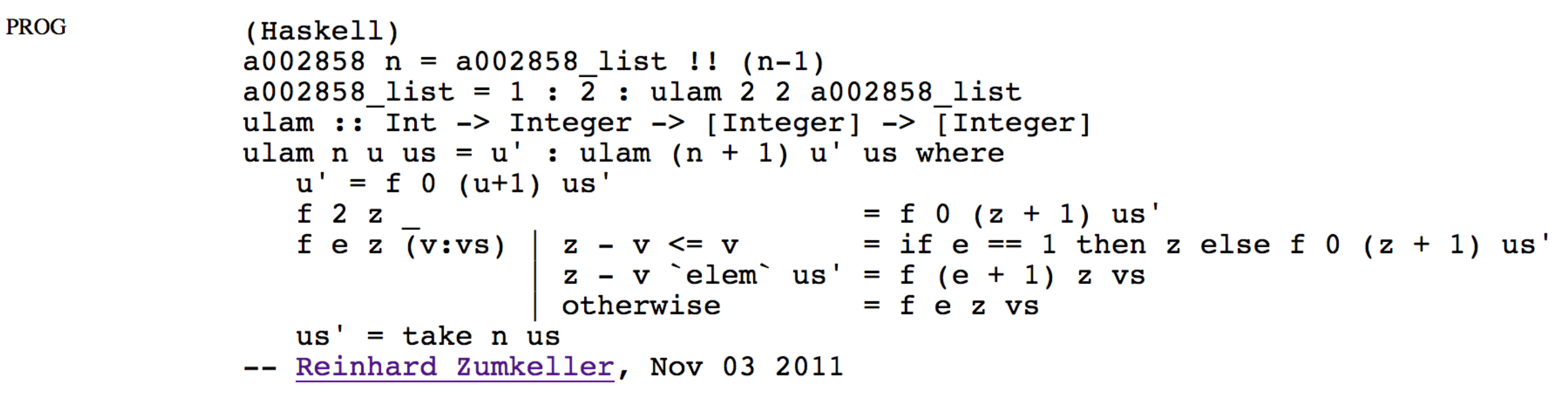}
\end{tcolorbox}
\vspace{-4mm}
\caption{Filter-based program from OEIS that computes Ulam numbers~\cite{ulam-oeis}.}
\label{fig:ulam-filter-OEIS}
\end{figure}

\myParagraph{\hermione:}
I found on the OEIS page \cite{ulam-oeis} 
another Haskell program computing the Ulam sequence 
attributed to Reinhard Zumkeller (Fig.~\ref{fig:ulam-filter-OEIS}). 
This program follows the filtering 
approach you have hinted to us for writing the C function.

The core of this program is the function \code{f}: 
it has three arguments, \code{e}, \code{z}, and a tail \code{v} 
of the finite list of already computed Ulam numbers. 
Function \code{f} checks if the Ulam candidate \code{z} is the next Ulam number. 
To do this, the first parameter \code{e} is used to count how many distinct 
sums of terms in \code{v} 
give as result \code{z}.
If this number is 1 it returns \code{z} as the next Ulam number, 
otherwise \code{f} calls itself recursively to verify if \code{z+1} is the next Ulam 
number.
To save computations, as soon as the first parameter \code{e} becomes 2,  
it discards \code{z} and it starts to consider \code{z+1} as the next Ulam candidate.
Function \code{f} finds a new sum giving \code{z} as result, by checking 
if \code{z-v} belongs to \code{us'} (calling the standard Prelude function \code{elem}). 
The list \code{us'} another copy (not local to \code{f}) of 
the list of already computed Ulam numbers. 

The function \code{ulam} is the engine that produces Ulam numbers. 
It takes three arguments, \code{n}, \code{u}, \code{us}, 
which are, respectively, the number of elements of the sequence already known, 
the last computed Ulam number (that is needed to start the computation of \code{f}),  
and an (infinite) list whose first \code{n} elements are the first \code{n} Ulam numbers.

At first sight, too many copies of already computed Ulam numbers!

\subsection{Generating Ulam Numbers} 

\myParagraph{\snape:}
On the basis of what we learned from the Hamming problem,  
it is worth investigating whether a generative solution   
to the problem of computing Ulam numbers could be more effective than a filtering one, 
even though filtering will be always part of any Ulam numbers computation. 

Things are interesting here, because we are in a different scenario, now. 
Even though Ulam himself conjectured that the density of this sequence is 0, 
experimentally we know that its density converges to $0.074$, that is 
$u_n$ tends to be $13.51n$.

\myParagraph{\hermione:} I've just done something! 
Getting inspiration from my original generative Hamming solution (Fig.~\ref{fig:composti-generative-2}), 
I wrote a program that selects the next Ulam number, let's say $u_{n+1}$ 
from a list $c = [c_1, \ldots, c_n]$ of $n$ infinite lists of candidates, 
where each $c_i$ is an infinite suffix of $[u_i+u_{i+1}, u_i+u_{i+2},\ldots]$. 
My program (Fig.~\ref{fig:ulam-generative-W}): 
\begin{enumerate*}[label=({\arabic*})] 
\item computes the list \code{chs} of heads of lists in \code{c},
\item computes the minimum \code{m} in \code{chs}, 
\item computes the number \code{k} of occurrences of \code{m} in \code{chs}. 
\item then, if \code{k} is strictly greater than 1, it simply 
calls recursively \code{nextU}, on the list of candidates \code{c'} obtained from \code{c} 
by removing occurrences of \code{m} 
(that appears as the head of some lists);
\item 
otherwise, if \code{k} is 1, \code{m} is the new Ulam number and a new list 
$c_{n+1} = [u_{n+1}+u_{i+2}, u_{n+1}+u_{n+3},\ldots]$ is added to \code{c'}\ldots 
\end{enumerate*}
({\em \hermione\ has just pushed a button on her laptop, probably to run her program, and now she is staring at the screen})
Unfortunately... 
    
\myParagraph{\snape:} \ldots It fails to compute $u_4=6$, I suppose. Because your program tries to get the head of a  
not already computed list. Your informal argument to prove that no productivity issues arise in your solution to the Hamming 
problem does not work here. Your Ulam program can repeat head extractions more than once, before computing a new number.  

\myParagraph{\hermione:} Yes\ldots But we need only sums of pairs among $u_1, u_2, \ldots, u_n$, so I can add to each list 
a number\ldots

\begin{figure}
\begin{center}
\begin{minipage}{0.72\linewidth}
\begin{tcolorbox}[colback=white, boxrule=0pt, toprule=1pt, bottomrule=1pt, left=0pt]
{\small
\begin{verbatim}
  nextU :: (Ord a, Num a) => [a] -> [[a]] -> [a]

  nextU u@(hu:tu) c = if k>1 then nextU u c' 
                             else m:nextU tu c'' 
     where       
       chs = map head c    
       m = foldr1 min chs            
       k = foldl1 (+)(map (\x->if x==m then 1 else 0) chs)
       c' = map (\x@(hx:tx)->if hx==m then tx else x) c
       c'' = map (hu+) tu:c'

  ulam = 1:2:nextU tu [map (1+) tu] where tu = tail ulam
\end{verbatim}
}
\end{tcolorbox}
\end{minipage}
\end{center}
\vspace{-4mm}
\caption{Generative solution to compute Ulam numbers. Wrong solution (productivity problems).}
\label{fig:ulam-generative-W}
\vspace{-2mm}
\end{figure}

\myParagraph{\snape:} Of course, \mrsGranger. I'm pretty sure that you can quickly arrange your code and make it work. 
But your program will be even murkier than it is now! 

What we really need to compute $u_{n+1}$ are the increasing finite list of all sums $u_i+u_j$, for $1\leq i<j\leq n$. 

\myParagraph{\hermione}
That's right: it is enough, whenever a new Ulam number is generated, 
to generate the finite list of sums with {\em its predecessors}!

The basic idea of my new generative program (Fig.~\ref{fig:ulam-generative}) 
is the following: whenever $u_n$ has been computed, it generates all sums 
$u_1+u_n, u_2+u_n,\ldots,u_{n-1}+u_n$ and it inserts them in a list of 
Ulam candidates. I have defined a sum type \code{UlamC} 
to distinguish numbers that have been obtained as the sum of just one pair  
of earlier terms (\code{Unique n}) from those that are sum of more than one pair (\code{Duplic n}).
%

Each recursive call of the engine \code{nextU} assumes that the first element of 
the list of candidates is the next Ulam number. 
To guarantee this precondition:   
\begin{enumerate*}[label=({\arabic*})] 
\item we insert all sums generated by the new Ulam number with its predecessors \code{preds} 
(selected by \code{takeWhile (<u) ulam}) into the tail \code{tc} of the list \code{c} of candidates;
\item function \code{insert} takes care to correctly classify new sums into \code{Unique} and \code{Duplic} and keep them 
in increasing order;  
\item finally, the prefix of duplicated sums is removed from the list of candidates by 
\code{dropWhile (isNotUlam) c'} to ensure that \code{c''} satisfies preconditions of \code{nextU} in the \code{nextU} recursive call.
\end{enumerate*}


Unfortunately, we can not remove a number from the list of Ulam candidates 
as soon as we find at least two distinct pairs that give it as their sum, 
simply because this number could be generated again as the sum of a pair containing an 
Ulam number that will be computed later.

\myParagraph{\snape:}
Not bad, \mrsGranger: 
the main benefit of your program, compared to any program based on a 
filtering approach, is to avoid checking numbers that are not the sum of any pair of 
Ulam numbers, i.e. all numbers that belong to the above mentioned sequence $v$ (see  
Section \ref{sec:ulamNumbers}). 
Its computational complexity is clearly quadratic in $n$ (to compute $u_n$)
as it computes $n(n-1)/2$. Experimentally 
it performs much better than 
the Reinhard Zumkeller program of Fig.~\ref{fig:ulam-filter-OEIS}
(see Section \ref{sec:complexity} later on). 

\begin{figure}[t]	
\begin{center}
\begin{minipage}{0.75\linewidth}
\begin{tcolorbox}[colback=white, boxrule=0pt, toprule=1pt, bottomrule=1pt, left=0pt]
{\small
\begin{verbatim}
  data UlamC = Unique Int | Duplic Int

  value (Duplic x) = x
  value (Unique x) = x

  isNotUlam (Duplic x) = True
  isNotUlam (Unique x) = False

  insert :: [Int] -> [UlamC] -> [UlamC]
  insert  s [] = map (\x->(Unique x)) s
  insert []  c = c
  insert s@(hs:ts) c@(hc:tc) 
      | hs <  m    = (Unique hs):insert ts c
      | hs == m    = (Duplic hs):insert ts tc
      | otherwise  = hc:insert s tc
    where m = value hc

  nextU :: [UlamC] -> [Int] -> [Int]
  nextU ((Unique u):tc) ulam = u:nextU c'' ulam where
      preds = takeWhile (<u) ulam
      c'    = insert (map (+u) preds) tc
      c''   = dropWhile isNotUlam c'

  ulam = 1:2:nextU [(Unique 3)] ulam 
\end{verbatim}
}
\end{tcolorbox}
\end{minipage}
\end{center}
\vspace{-4mm}
\caption{Generative solution to the computation of Ulam numbers.}
\label{fig:ulam-generative}
\vspace{-2mm}
\end{figure}

\subsection{Back to the Filtering Approach}

\myParagraph{\snape:}
As I told you several times, writing programs in different languages can give you 
new insights into solving a problem.
When I assigned to you the problem of writing a C function 
that computes the Ulam sequence, I had in mind essentially a filtering approach. 
And almost all of you followed this path of reasoning 
(perhaps influenced by my implicit or explicit hints in classroom 
problem discussions).

Therefore, after writing the Haskell generative program 
in Fig.~\ref{fig:ulam-generative}, it has been natural for me to implement 
its corresponding C program and to check if the generative C  
program is faster than the C program based on a filtering approach. 

\myParagraph{\hermione:}
I did it too, and the answer is no. My filtering based C program 
(see Appendix, Fig.~\ref{fig:c-filtrative}) runs 
significantly faster than my generative C program (see Appendix, Fig.~\ref{fig:c-generative}).
For a while, I interpreted this surprising fact as a nice property of the 
lazy computational mechanism behind Haskell that automatically avoids computing  
not needed sums. 
I have spent the last week trying to improve my generative C program 
in order to get a program that carefully computes only needed sums. 
At the end of the week, I gave up.

\myParagraph{\snape:}
Unfortunately for you, 
the moral is not that your generative C program is not good enough, 
but rather than the OEIS Haskell filtering program is not good enough!

Indeed, upon closer inspection, my filtering C program 
(and I suppose your filtering C program too) is essentially based on 
a slightly refined idea, that came natural in my mind 
the first time I designed a C program to compute Ulam numbers:  
let us suppose $z$ be a candidate to be $u_{n+1}$ 
and that we are examining $u_i$ and $u_j$ ($1\leq i<j\leq n$).    
If $u_i+u_j<z$, then we should continue by checking if $u_{i+1}+u_j>u_i+u_j$ 
is equal to $z$. 
Symmetrically, if $u_i+u_j>z$, then we should continue by checking 
if $u_{i}+u_{j-1}<u_i+u_j$ is equal to $z$. Finally, if $z=u_i+u_j$ and this is the first sum that gives $z$ as result, we check if there exist other sums, 
starting from $u_{i+1}+u_{j-1}$. 
If $u_i+u_j$ is not the first sum encountered so far giving $z$ as result, 
we can immediately conclude that $z$ is 
not $u_{n+1}$. 
If $j\geq i$ and if $z$ has been obtained just once, then we can conclude that $z$ is $u_{n+1}$, otherwise (that is, if no sum has been found giving $z$ as a result), 
we can again conclude that $z$ is not $u_{n+1}$. 
Starting this process with $i=1$ and $j=n$, 
we can efficiently (even though still linear in $n$) verify whether $z$ is $u_{n+1}$ or not. 

To implement this idea, we need to scan the sequence $u_1,\ldots,u_n$ 
in both directions, and not without reason, I required to you, in your C program, 
to store Ulam numbers in a double-linked list, in the hope that 
you would have exploited the possibility of scanning a double linked list in both directions (using arrays, this idea works  almost the same).  
\smallskip

Now, our key question is finally clear: how to efficiently implement this idea when  working with Haskell lists, which can be visited only forwards? 

\begin{figure}	
\begin{center}
\begin{minipage}{0.68\linewidth}
\begin{tcolorbox}[colback=white, boxrule=0pt, toprule=1pt, bottomrule=1pt, left=0pt]
{\small
\begin{verbatim}
  reverseK k l = revKAux k l [] where 
      revKAux 0 l m = m
      revKAux k (h:t) m = revKAux (k-1) t (h:m) 

  howManySums n  []  _   = 0
  howManySums n   _  []  = 0
  howManySums n u@(hu:tu) r@(hr:tr)
     | hr <= hu = 0
     | hu+hr == n = 1+howManySums n tu tr
     | hu+hr  < n = howManySums n tu r
     | otherwise  = howManySums n  u tr

  nextU n k l = 
      if howManySums n l (reverseK k l)==1
         then n:nextU (n+1) (k+1) l
         else nextU (n+1) k l

  ulam = 1:2:nextU 3 2 ulam       
\end{verbatim}
}
\end{tcolorbox}
\end{minipage}
\end{center}
\vspace{-4mm}
\caption{Na\"ive filtering program.}
\label{fig:ulam-filter-naive}
\vspace{-2mm}
\end{figure}

\myParagraph{\hermione:}
I've got it: we need both Ulam numbers and the  
list of Ulam numbers computed so far in the reversed order. 
Referring to your analysis, incrementing $i$ means advance on the Ulam sequence, decrementing $j$ means advance on the reversed Ulam sequence, and finally, since $u$ is an increasing sequence, the condition $i<j$ is equivalent to the condition $u_i<u_j$,   
and therefore we have to continue until the head of the Ulam sequence is strictly less than 
the head of its reversed counterpart. 

I have made a first attempt to do this (Fig.~\ref{fig:ulam-filter-naive}).
This program is much faster than the OEIS program (Fig.~\ref{fig:ulam-filter-OEIS}), 
but it still performs rather worse than the generative version (Fig.~\ref{fig:ulam-generative}).

\begin{figure}	
\begin{center}
\begin{minipage}{0.68\linewidth}
\begin{tcolorbox}[colback=white, boxrule=0pt, toprule=1pt, bottomrule=1pt, left=0pt]
{\small
\begin{verbatim}
  nextUlam :: (Num a, Ord a) => a -> [a] -> [a] -> [a]
  nextUlam n u uR
    | isUlam n 0 u uR = n:nextUlam (n+1) u (n:uR)
    | otherwise       = nextUlam (n+1) u uR 
   where
     isUlam n 2 _ _ = False
     isUlam n h u@(hu:tu) r@(hr:tr) 
       | hr<=hu    = h == 1
       | hr+hu==n  = isUlam n (h+1) tu tr
       | hr+hu<n   = isUlam n   h   tu r
       | otherwise = isUlam n   h   u  tr

  ulam = 1:2:nextUlam 3 ulam [2,1]
\end{verbatim}
}
\end{tcolorbox}
\end{minipage}
\end{center}
\vspace{-4mm}
\caption{Optimised filtering solution.}
\label{fig:ulam-filter-final}
\vspace{-2mm}
\end{figure}

\myParagraph{\snape:}
Your program has essentially two problems: 
\begin{enumerate*}[label=({\arabic*})] 
\item
function \code{howManySums n l m} always computes all possible sums (without stopping when it finds the second one equal to \code{n}) and, more importantly,  
\item 
it reverses the list of Ulam numbers computed so far, at each invocation of 
function \code{nextU}, that is each time it starts to examine if a new candidate is 
the next Ulam number.
\end{enumerate*}

We can solve both these problems by using standard tricks in functional programming
that can be summarised in the slogan that was often repeated by my Maestro, 
prof. Corrado B\"ohm: 
``functional programming is stateless, but function parameters are stateful'', meaning that 
parameters can be used to store intermediate results of some computations.

\myParagraph{\hermione:} Oh right... and we can use the same trick in Zumkeller's function  \code{f}  
(Fig.~\ref{fig:ulam-filter-OEIS}) and the trick in the \code{reverseK} function (Fig.~\ref{fig:ulam-filter-naive}). 
Function \code{howManySums} can be defined with a supplementary parameter! 
Anyway, I have written a new boolean function \code{isUlam n h u r} that uses the parameter \code{h} to transmit information to its recursive calls on how many sums equal to \code{n} have been found so far, and immediately returns \code{False} as soon as the value of \code{h} is 2. 

We can avoid reversing the Ulam sequence at each invocation of function \code{nextU} 
just by adding a supplementary parameter to it that keeps the reversed list of Ulam numbers  computed so far. This, in turn, makes the parameter \code{k} of \code{nextU} old version useless, as it was needed 
in the program of Fig.~\ref{fig:ulam-filter-naive} to reverse just the first   
\code{k} elements of the potentially infinite list of Ulam numbers. 

Function \code{nextU n u r} now has three parameters: the next candidate to be checked, the Ulam sequence under construction, and the finite reversed list of Ulam numbers computed so far. 
The precondition that \code{r} is the reverse of \code{u} is guaranteed by the first call and then it is maintained since whenever \code{nextU} produces a new Ulam number, 
we add it in front of \code{r}. 

This program is also as short as the OEIS program and, in my opinion, even cleaner!

\myParagraph{\snape:} Excellent work, \mrsGranger. To reward your brilliant contributions to our class discussion in these lectures about integer sequences, I want you to prepare some experimental results and some discussion about computational complexity of all Ulam programs 
discussed so far. Next Monday, for the last lecture, you will give a classroom presentation. 

\myParagraph{\hermione\ (screaming):} But \snape! It's Friday afternoon... 

\myParagraph{\snape:} Yes, I know. So, \mrsGranger, the moral is: 
don't was-te ti-me scream-ing!

\subsection{Complexity Considerations and Experiments}
\label{sec:complexity}

\begin{figure}
\centering
\vspace{-3mm}
\includegraphics[scale=0.53]{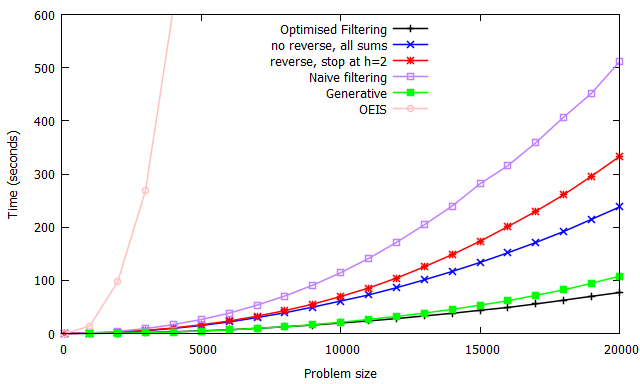}
\vspace{-3mm}
\caption{Program running time (up to $n=20.000$)}  
\label{fig:g1}
\vspace{-4mm}
\end{figure}

\myParagraph{\hermione:}
It is known that the computation of $u_{n+1}$ having $u_1,\ldots, u_n$   
costs at least $\mathcal{O}(n)$. Since computing $u_{n}$ requires to know 
all its predecessors,  
it takes $\mathcal{O}(n^2)$ to compute the first $n$ terms of the sequence. 
Therefore, as observed in 
\cite{gibbsMcCranie}, 
any ``direct'' algorithm is at best quadratic in $n$. 
This is true, only if one properly takes advantage of 
storing Ulam numbers in ordered lists, as it is natural. 
All our programs do this, and hence they are all quadratic. 

By contrast, to check if a candidate $z$ is $u_{n+1}$,
the OEIS program calls,    
for all $i\leq n$, the linear function \code{elem} to verify if $z-u_i$ is one of $u_1,\ldots, u_n$. 
Therefore, it computes $u_{n+1}$ in $\mathcal{O}(n^2)$ and 
the first $n$ Ulam numbers in $\mathcal{O}(n^3)$.
This is confirmed by my experimental results (see Fig.~\ref{fig:g1}).

I run all programs 
by using the compiler GHC 8.2.1 on my laptop, 
TOSHIBA SATELLITE Z30-B-100, Windows 8.1, CPU Intel\textregistered Core$^\mathrm{TM}$ x64, i-5-5200U, 
2.20GHz, 8GB of RAM. 
Here I have plotted running times of all programs 
we considered in the last two lectures 
up to $n=20.000$.

\myParagraph{\snape:} Strictly speaking, this is not true, \mrsGranger. I don't see in your picture 
any mention to the running time of the 67 characters program!

\myParagraph{\hermione:} But\ldots \snape, the first experiment with $n=1000$ is 
still running after 3 days!

\myParagraph{\snape:} You could plot some dots in the up-left corner of your picture, 
but please, go on.

\myParagraph{\hermione:} Besides na\"ive filtering, generative, 
and optimised filtering, I have considered also two more programs, 
each of which makes use of just one of our two optimisations leading from 
na\"ive filtering to its optimised version: one avoids  
reversing the list of Ulam numbers at each \code{nextU} invocation 
(``no reverse, all sums''), 
and the other one 
discards the candidate at hand as soon as it finds two sums of 
earlier terms equal to it 
(``reverse, stops at $h=2$''). 
Interestingly, in Fig.~\ref{fig:g1}, we can see that the generative program 
looks faster than both of these two ``partially optimised'' programs. 

To refine our theoretical analysis, we need to get rid of $\mathcal{O}$ notation 
and to include constants multiplying $n^2$ into play. 
This is a risky game, in the sense that now we must be as careful as possible
to estimate these constants.
We assume the following experimental evidence: $u_n$ is 
approximately $13.5n$ and up to $u_n$ we have about $2.5n$ 
numbers that are not sums of two Ulam numbers.
%
%
%

The na\"ive filtering program computes $u_{n+1}$ 
checking (on average) $13.5$ candidates. For each of them, 
it invokes functions \code{howManySum} 
and \code{reverseK} and both of them cost $n$. 
Applying the Gauss formula for the sum of the first $n$ naturals, 
we get that computing $u_n$ costs $13.5\cdot 2\sum_{i=0}^n i$, that is 
$13.5n^2$.
Avoiding \code{reverseK}, the cost decreases to 
$6.75n^2$.
It is not clear to me how to estimate the impact of stopping at $h=2$ 
but it is really relevant, as it is clear from experimental results.   

As for the generative program, the computation of $u_{n+1}$ requires 
\begin{enumerate*}[label=({\arabic*})]
\item to discard prefix of duplicate sums in the candidate list (this can be ignored, 
since it does not depend on $n$, requiring on average $11\ (=13.5-2.5)$ steps), 
\item
to compute sums of the new Ulam number with its predecessors that costs $n$, and 
\item
inserting such sums into the list of candidates: function \code{insert} is a sort of \code{merge} 
between a list of numbers in the interval $[u_{n}+1,u_n+u_{n-1}]\!\approx\! [13.5n,27n]$, but again without numbers that are not sums of Ulam numbers, and  a list of $n$ 
numbers.
Therefore, \code{insert} costs about $12n$.
\end{enumerate*}
Putting everything together as before with Gauss formula, we have about $6.5n^2$.

The running time of the Haskell generative program is very close to that of the optimised Haskell filtering program. 
Interestingly, the C program based on filtering is more than 6 times faster than C generative program for $n=20.000$.
     
  
\myParagraph{\studenteUno:} Excuse me, \hermione, I have many questions for you: 
\begin{enumerate*}[label=({\arabic*})]
\item
Only extraction of the next Ulam number forces the list of candidates to be really computed  
by the lazy evaluation of Haskell. How this come into play? This could be related to your observations 
about comparison between corresponding C programs.
\item
Long time ago, you showed me a paper about the Sieve of Eratosthenes \cite{sieve09}: there, advanced data structures were used instead of lists. Could these data structures improve running time of your programs?
\item Could we think about a hybrid approach? 
\end{enumerate*}

\myParagraph{\snape} Prime numbers, Sieve of Eratosthenes, making use of advanced data structures, complexity analysis of lazy programs\ldots~exciting topics for holiday homework assignments! Excellent work, \mrsGranger.
We stop here: for this year, it's enough. 
 


\section{Epilogue}
\myParagraph{\snape\ (farewell speech):}
It has been a great pleasure for me 
to have all of you as students  
in my course of {\sl ``Functional and Imperative Programming Pearls''}. 
Working together, we learned several lessons:


\begin{itemize}[wide] 
\item
reading old books, especially when written by inspired authors, 
can transmit new ideas;
\item
asymptotic analysis of computational complexity is a solid 
foundation that you must never forget. Nevertheless, 
writing good and effective programs sometimes requires some ingenuity, 
even if you are implementing the best available algorithm to solve 
a given problem;
\item 
Programming Languages are also conceptual tools:
writing (and thinking) programs 
in different languages 
may help us to stay away from artless results. 
As an example, 
getting inspiration from functional programming, 
you can write C code as clean as possible. 
Similarly, some code optimisations could be more easily figured out   
when thinking a program as a sequence of memory transformations:  
then you can find how to apply such ideas in your functional code;
\item 
Beauty is important: often (unfortunately not always) 
a small and elegant solution demonstrates  
that you have grasped the core of a problem, 
instead of loosing yourself in analysing useless corner cases;
\item
it is important to write programs according to language's etiquette or idiom: 
this helps you to quickly understand 
code written by other smart programmers and (possibly) collaborate with them. 
\end{itemize}

And of course, don't forget your holiday homework assignments!

\subsection*{Acknowledgment}
We wish to thank anonymous reviewers of an earlier version of this paper 
for their useful comments and suggestions.

\bibliographystyle{alpha}
\bibliography{biblio}
\newpage
\section{Afterword}
\myParagraph{\hermione, \studenteUno, and \studenteDue\ are together at the end of summer.}

\myParagraph{\hermione:} You look so blissful, guys. What about holiday homework assignments? 
I'm struggling with an exercise given by \snape.  
I do not understand how to compute prime numbers taking advantage 
of my nice program that computes composites\ldots

\myParagraph{\studenteUno:} I found something similar in \cite{sieve09}: in the Epilogue there is 
a sharp program credited to Richard S. Bird. 

\myParagraph{\hermione:} Yes, I read that paper, but that program works with 
{\em multiples} of primes: I would like to program  
the Sieve of Eratosthenes in such a way that each composite is crossed over just once!
I'm trying to work with a list whose first element is the list 
of composites of two, the second element is the list of numbers that are composite 
of three but not of two, and so on... and then complement the merge of such lists, 
probably applying techniques in \cite{sieve09} and \cite{Bird15}.

\myParagraph{\studenteDue:} Yes, but doing so, one need to complement several times infinite lists. 
I did some work in this direction as you,  
for example, have a look to this program (Fig.~\ref{fig:sieve1}).
I wrote also some optimised versions, but 
while the code gets more and more obscure, improvements 
are very little\ldots  
trial division performs always much better than my sieves! 
My laptop got hot and 
\code{GCHi} got crazy because of garbage collector!

\begin{figure}[h]	
\begin{center}
\begin{minipage}{0.68\linewidth}
\begin{tcolorbox}[colback=white, boxrule=0pt, toprule=1pt, bottomrule=1pt, left=0pt]
{\small
\begin{verbatim}
   minus (x:xs) (y:ys)
      | x==y = minus xs ys
      | x<y  = x:minus xs (y:ys)
      | x>y  = minus (x:xs) ys

   mySieve (x:xs) = x:mySieve (minus xs (map (x*) (x:xs)))

   primes = mySieve [2..]
\end{verbatim}
}
\end{tcolorbox}
\end{minipage}
\end{center}
\vspace{-4mm}
\caption{Sieve of Eratosthene I. Inefficient version.}
\label{fig:sieve1}
\vspace{-2mm}
\end{figure}
 
\myParagraph{\hermione:} You are right! It is not trivial to generate all composites of primes {\em without complementing} infinite lists, as your program does by using function \code{minus}.

Maybe that a new approach to the Hamming problem could help (Fig.~\ref{fig:hammingRevisited}): 
the idea is that composites of $x_1, \ldots x_n$ are the powers $P=\{x_1, x_1^2, x_1^3,\ldots\}$ of $x_1$, 
plus all the composites $C$ of $\{x_2, \ldots, x_n\}$, plus all the products of elements in $P$ with elements in  
$C$, that is the set $\{p\cdot c~|~ p\in P, c\in C\}$.

\myParagraph{\studenteUno:} Since Haskell allow one to mimic set comprehension via list comprehension, 
why don't you define \code{allProducts} simply as: 
\begin{center}
\code{allProducts xs zs = [x*z | x<-xs, z<-zs]}
\end{center}

\myParagraph{\hermione:}
You are right, but again, due to productivity problems, this does not work in the solution of this problem.  
I take inspiration from the Sieve of Eratosthenes of prof. Bird in \cite{sieve09}: 
we know that the first element of \code{map (x*) z} is {\em always} smaller than 
the first element of \code{allProducts xs z} and I take advantage of this fact to {\em compute} a new 
product even when products computed by recursive calls are not ready. 

To do so, I borrow the \code{merge'} function of prof. Bird, that always take the first element from 
the first parameter (here I call it \code{unionP}, and \code{union} in its definition is the function defined in 
Fig.~\ref{fig:composti-generative-union}, with standard equations for empty lists). 
Having this idea in mind, we observe that prime numbers satisfy this recursive equation: $primes = \mathbb{N}_{\geq 2} \setminus (composites ~ primes)$, that in turn is equivalent to:

\begin{figure}	
\begin{center}
\begin{minipage}{0.75\linewidth}
\begin{tcolorbox}[colback=white, boxrule=0pt, toprule=1pt, bottomrule=1pt, left=0pt]
{\small
\begin{verbatim}
   unionP (x:xs) ys = x:union xs ys

   allProducts x [] = []
   allProducts [] z = z
   allProducts (x:xs) z =  map (x*) z `unionP` allProducts xs z

   hammingAux [] = []
   hammingAux (x:xs) = p `unionP` h `unionP` allProducts p h where 
       p = x : map (*x) p
       h = hammingAux xs
   hamming l = 1:hammingAux l
\end{verbatim}
}
\end{tcolorbox}
\end{minipage}
\end{center}
\vspace{-4mm}
\caption{Hamming problem, revisited.}
\label{fig:hammingRevisited}
\vspace{-2mm}
\end{figure}

\begin{figure}[b]	
\begin{center}
\begin{minipage}{0.75\linewidth}
\begin{tcolorbox}[colback=white, boxrule=0pt, toprule=1pt, bottomrule=1pt, left=0pt]
{\small
\begin{verbatim}
   allProducts (x:xs) z = map (x*) z `unionP` allProducts xs z

   composites (x:xs) = 
          tail p `unionP` allProducts p (union xs c) `unionP` c 
     where
        p = x : map (*x) p
        c = composites xs

   primes = 2:([3..] `minus` composites primes)\end{verbatim}
}
\end{tcolorbox}
\end{minipage}
\end{center}
\vspace{-4mm}
\caption{Sieve of Eratosthene II. Efficient version.}
\label{fig:sieve2}
\vspace{-2mm}
\end{figure}
%
\begin{center}
$primes = \{2\}\cup \mathbb{N}_{\geq 3} \setminus (composites ~ primes)$ 
\end{center}
useful to start the lazy computation of the sequence of primes. This is the main idea of my efficient prime numbers generator 
(Fig.~\ref{fig:sieve2}), 
that generates {\em just once} each composite number.   
Finally, we have to take care to avoid to generate prime numbers in function \code{composites}, but to use them in new products! 
Therefore, the function \code{composites} is slightly more complicated than the function \code{hammingAux} 
in the solution of the Hamming problem: it removes $x$ from the sequence of powers of $x$, but it insert prime numbers 
greater than $x$ in the set of composites of successors of $x$ to multiply them with powers of $x$. 

\myParagraph{\studenteDue:} Nice work, but\ldots what about its performance?

\myParagraph{\hermione:} Rather good. It is about twice faster than Bird's program to compute the first 100.000 primes 
and almost three times faster to compute the first 500.000 primes. Therefore, I conjecture that it is asymptotically faster. 

I should calculate its complexity carefully, but now, I'm really tired. After all, tomorrow is another day.

%
%
%

\appendix
\newpage
\section{C programs}
For the sake of completeness, in this section we show C programs for computing Ulam numbers 
we refer to in the paper text. 
In Fig.~\ref{fig:c-filtrative}, we show C implementation of function \code{nextU} based
on filtering. It implements the same algorithm as the Haskell program in Fig.~\ref{fig:ulam-filter-final}.

Ulam numbers are stored in a double linked list. The type \code{listDL} 
allow us to scan a list in both direction (using pointers \code{next} and \code{previous}) and 
to have direct access to the first and the last node of a list (using pointers \code{first} and \code{last} 
in the list descriptor).

In Fig.~\ref{fig:c-generative}, we show a C implementation of  
the generative approach of the Haskell program in Fig.~\ref{fig:ulam-generative}. 
Here, we use an array to store Ulam numbers. Since direct access is not important in 
these computations and (probably) thanks to optimised memory allocation implemented in the C run-time 
environment, experimentally, using arrays speeds-up running time just about $5\%$ with respect to list-based implementations.

The functions \code{insert} mimics the Haskell \code{insert} function, and function \code{elim} implements 
the elimination of prefixes of non-Ulam numbers in the list of candidates performed by the line \code{dropWhile isNotUlam c'} in 
the Haskell program.
\bigskip

\begin{figure}[h]
\begin{center}
\begin{minipage}{0.68\linewidth}
\begin{tcolorbox}[colback=white, boxrule=0pt, toprule=1pt, bottomrule=1pt, left=0pt]
{\small
\begin{verbatim}
   int nextU(listDL u){
      listDL inf; 
      listDL sup;
      int new = u->last->val;
      do { new++;
           inf = u->first;
           sup = u->last;
           int nt = 0;
           while (nt<2 && inf->val<sup->val){
              int v=inf->val + sup->val;
              if (v==new){
                  nt++;
                  inf=inf->next;
                  sup=sup->prev;
              } else if (v>new) sup=sup->prev;
              else inf=inf->next;
           }	
      } while (nt!=1);
      addLast(new, u);
      return new;	
    } 
\end{verbatim}
}
\end{tcolorbox}
\end{minipage}
\end{center}
\vspace{-4mm}
\caption{Filtering C program computing Ulam numbers, implemented with double linked lists.}
\label{fig:c-filtrative}
\vspace{-2mm}
\end{figure}

\begin{figure}[t]
\begin{center}
\begin{minipage}{0.68\linewidth}
\begin{tcolorbox}[colback=white, boxrule=0pt, toprule=1pt, bottomrule=1pt, left=0pt]
{\small
\begin{verbatim}
  typedef unsigned long int bigint;

  list elim(list c){
      if (!c || c->occ==1) return c;
      c=tail(c);
      return elim(c);
    } 

  list insert(list c, bigint u[], int k, int j){
      bigint v = u[k]+u[j];
      if (j==k) return c;
      if (!c || v<c->val) 
         return cons(v,insert(c,u,k,j+1));
      if (c->val < v) 
         c->next = insert(c->next,u,k,j);
      else {
         c->occ++;
         c->next = insert(c->next, u, k, j+1);
      }
    return c; 
  }

  void nextU(bigint u[], int k, list* c){
      list d = elim(*c);
      u[k] = d->val;
      d = tail(d);
      *c = insert(d, u, k, 0);
    }
\end{verbatim}
}
\end{tcolorbox}
\end{minipage}
\end{center}
\vspace{-4mm}
\caption{Generative C program computing Ulam numbers (stored in an array) and a simple list of pairs (number, number of occurrences) for Ulam candidates.}
\label{fig:c-generative}
\vspace{-2mm}
\end{figure}

\end{document}